\documentclass{PoS}
\usepackage[numbers]{natbib}

\title{Cosmological 21cm experiments: \\Searching for a needle in a haystack}

\ShortTitle{EoR experiments: the foregrounds}

\author{\speaker{Vibor Jeli{\'c}} \thanks{on behalf of the LOFAR-EoR team}\\
        ASTRON, P.O. Box 2, 7990 AA Dwingeloo, the Netherlands\\
        E-mail: \email{jelic@astron.nl}}

\abstract{There are several planned and ongoing experiments designed to explore the Epoch of Reionization (EoR), the pivotal period during which the gas in the intergalactic medium went from being entirely neutral to almost entirely ionized. These experiments will probe the EoR, through the redshifted 21 cm line from neutral hydrogen, using radio arrays: e.g. Low Frequency Array (LOFAR) and Murchinson Widefield Array (MWA). Unfortunately however, the cosmological 21 cm signal is highly contaminated by astrophysical foregrounds and by non-astrophysical and instrumental effects. Therefore, to reliably detect the cosmological signal, it is essential to understand very well all data components, their influence on the desired signal and explore additional complementary or corroborating probes of the EoR. These proceedings give an overview of observational constrains of the foregrounds, present theoretical efforts to model the foregrounds, and discuss a problem of the foreground removal. The major results are presented for the LOFAR-EoR experiment.} 

\FullConference{ISKAF2010 Science Meeting - ISKAF2010\\
		June 10-14, 2010\\
		Assen, the Netherlands}

\begin{document}

\section{Introduction}
About four hundred million years after the Big Bang the first objects were formed, 
which then started to ionize the surrounding gas with their strong radiation. Six 
hundred million years later, the all-pervasive gas was transformed from a neutral 
to an ionized state. This pivotal period in the history of the Universe is called the 
Epoch of Reionization (EoR). It holds the key to structure formation and evolution, 
but also represents a missing piece of the puzzle in our current knowledge of the Universe.

\begin{figure}
\centering \includegraphics[width=1.\textwidth]{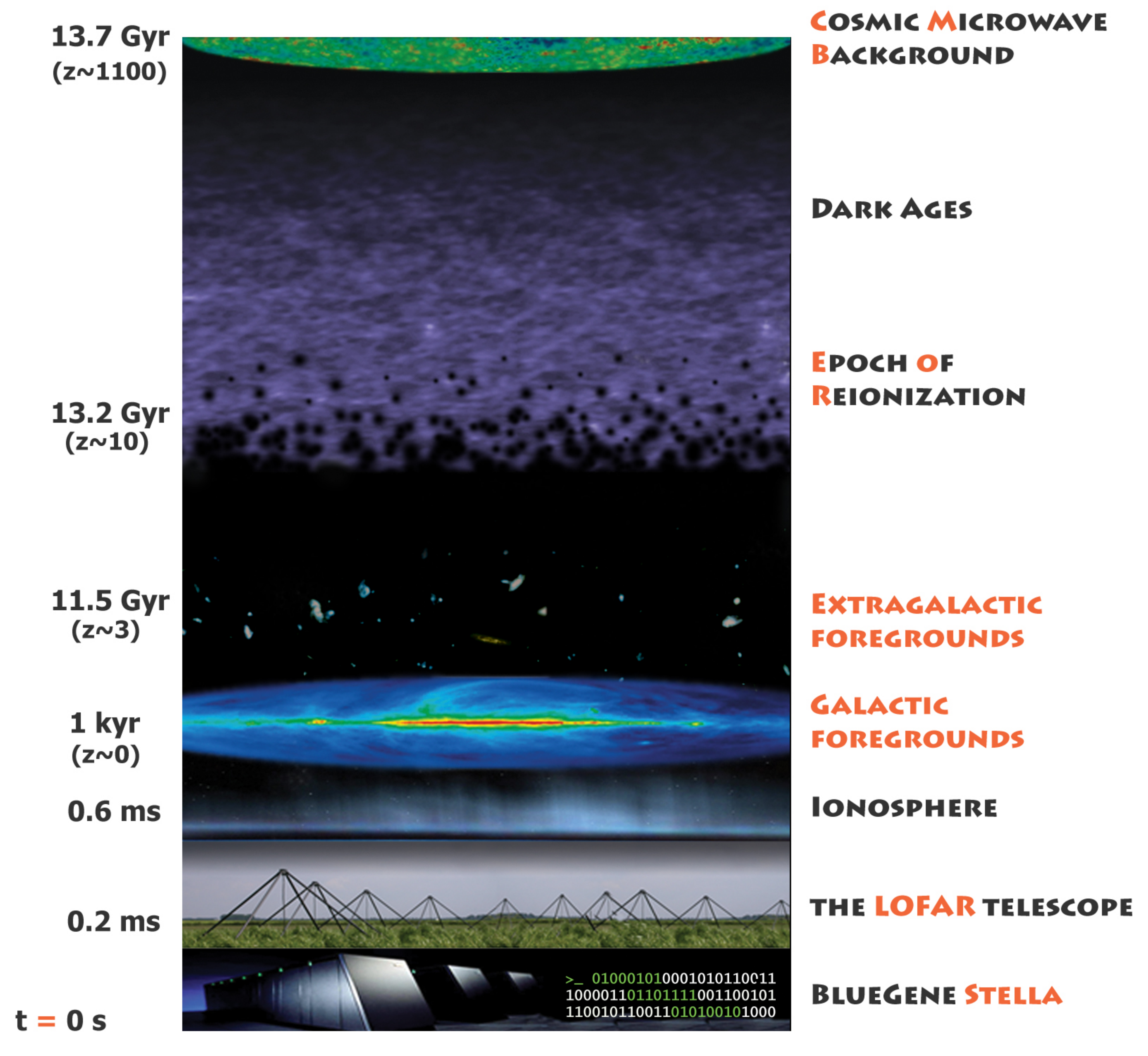}
\caption{This sketch illustrates all contributions to and contaminations of the observed signal in the case of the LOFAR-EoR key science project \cite{jelicPhD}. The former are: Galactic ($\sim 71\%$) and extragalactic ($\sim 27\%$) foregrounds \cite{jelic08,jelic10}, CMB $< 1\%$, and the EoR signal $\sim 0.01\%$ \cite{thomas09}; and the latter are: ionosphere, radio frequency interferences \cite{offringa10}, and instrumental effects and noise \cite{panos09}. On the left, a travel time of the observed signal is noted.}
\label{fig:EoRexp}
\end{figure}

Currently, this is changing through several planned and ongoing experiments 
designed to probe the Epoch of Reionization (EoR) by detecting redshifted 
21~cm emission line from neutral hydrogen: GMRT\footnote{Giant
Metrewave Telescope, http://gmrt.ncra.tifr.res.in}, LOFAR\footnote{Low
Frequency Array, http://www.lofar.org}, MWA\footnote{Murchinson
Widefield Array, http://www.mwatelescope.org/}, 21CMA\footnote{21
Centimeter Array, http://21cma.bao.ac.cn/}, PAPER\footnote{Precision
Array to Probe EoR, http://astro.berkeley.edu/$^\sim$dbacker/eor}, and
SKA\footnote{Square Kilometer Array, http://www.skatelescope.org/}.
However, all of these experiments are challenged by strong astrophysical foreground contamination, ionospheric distortions, complex instrumental response, and other different types of noise (see Fig.~\ref{fig:EoRexp}). 

In the frequency range of the EoR experiments ($\sim100-200~{\rm MHz}$) the foreground emission of our own Galaxy and extragalactic sources (radio galaxies and clusters) dominate the sky. In fact, the amplitude of this foreground emission is $4-5$ orders of magnitude stronger than the expected cosmological 21 cm signal. However, since the radio telescopes, which are used for the EoR observations, are interferometers, they measure only fluctuations of a signal. The ratio between the foregrounds and the cosmological signal is reduced to $2-3$ orders of magnitude.

In terms of physics, the foreground emission originates mostly from the interaction between relativistic charged particles and a magnetic field, i.e. synchrotron radiation. Galactic synchrotron radiation is the most prominent foreground emission and contributes about $70\%$  to the total emission at $150~{\rm MHz}$ \cite{shaver99}. The contribution from the extragalactic synchrotron radiation is $\sim 27\%$, while the smallest contribution ($\sim 1\%$) is from Galactic free-free emission, i.e. thermal radiation of an ionized gas. 

Given the prominent foreground emission that needs to be removed from the data in order to reliably detect the EoR signal, cosmological 21 cm experiments can be compared with \textit{finding the needle in the haystack} \cite{jelicPhD}. However, in the last decade there has been a slew of theoretical and observational efforts to explore and to understand all of the data components of the EoR experiments in order to prepare us for the real data. 

These proceedings give an overview of observational constrains of the foregrounds (Sec. 2), present  theoretical efforts to model the foregrounds (Sec. 3), and discuss a problem of the foreground removal (Sec. 4). The major results are presented for the LOFAR-EoR experiment, but these results are also applicable to the other EoR projects. The proceedings conclude with future perspective (Sec. 5).

\section{Observational constrains}
There are several all-sky maps of the total Galactic diffuse radio
emission at different frequencies and angular resolutions. 
The $150~{\rm MHz}$ map by \cite{landecker70} is the only all-sky map in the frequency range
($100-200~{\rm MHz}$) relevant for the EoR experiments, but has only
$5^\circ$ resolution.
 
In addition to current all-sky maps, a number of recent dedicated 
observations have given estimates of Galactic foregrounds in small selected areas.
\cite{ali08} have used 153 MHz observations with GMRT
to characterize the visibility correlation function of the
foregrounds. \cite{rogers08} have measured the spectral index of the
diffuse radio background between 100 and 200~{\rm MHz}. \cite{pen09}
have set an upper limit to the diffuse polarized Galactic emission; and
\cite[][see Sec.~\ref{sec:LFFE}]{bernardi09a,bernardi10}
obtained the most recent and comprehensive targeted observations 
with the Westerbork Synthesis Radio Telescope (WSRT).

The extragalactic foregrounds at the low radio frequencies are constrained
by the source counts from 3CRR catalog at 178 MHz \citep{laing83} and 
6C survey at 151 MHz \citep{hales88}. However, both catalogs are too shallow 
for the purpose of the EoR experiments.

\subsection{Galactic diffuse emission}
At high Galactic latitudes the minimum brightness temperature of the
Galactic diffuse emission is about $20~{\rm K}$ at $325~{\rm MHz}$
with variations of the order of 2 per cent on scales from 5 to 30
${\rm arcmin}$ across the sky \citep{debruyn98}. At the same Galactic
latitudes, the temperature spectral index of the Galactic emission is
about $-2.55$ at between 100 and 200 MHz \citep[e.g.][and references therein]{rogers08}
and steepens towards higher frequencies. Furthermore, the spectral index
gradually changes with position on the sky. This change appears to be
caused by a variation in the spectral index along the line of
sight. An appropriate standard deviation in the power law index, in
the frequency range 100--200~{\rm MHz} appears to be of the order of
$\sim 0.1$ \citep{shaver99}.

Using the obtained values at $325~{\rm MHz}$ and assuming the
frequency power law dependence, the Galactic diffuse emission is
expected to be $140~{\rm K}$ at $150~{\rm MHz}$, with $\sim 3~{\rm K}$
fluctuations. 

Studies of the Galactic polarized diffuse emission are done mostly at
high radio ($\sim 1 {\rm GHz}$) frequencies. At lower frequencies 
($\sim 350~{\rm MHz}$), there are several
fields done with the Westerbork telescope (WSRT). 
These studies \citep[][a review]{reich06} revealed a
large number of unusually shaped polarized small-scale structures of
the Galactic emission, which have no counterpart in the total
intensity. These structures are usually attributed to the Faraday
rotation effects along the line of sight. 

\subsection{LOFAR-EoR team observations with WSRT telescope}\label{sec:LFFE}
Recently, a comprehensive program was initiated by the LOFAR-EoR
collaboration to directly measure the properties of the Galactic radio
emission in the frequency range relevant for the EoR experiments. The
observations were carried out using the Low Frequency Front Ends
(LFFE) on the WSRT radio telescope. Three different fields were
observed. The first field was a highly polarized region known as the ``Fan
region'' in the 2nd Galactic quadrant  at a low Galactic latitude of $\sim10^\circ$
\cite{bernardi09a}. The second field 
was a very cold region in the Galactic halo ($l\sim170^\circ$) around the bright radio quasar
 3C196, and third was a region around the North Celestial Pole \citep[NCP, $l\sim125^\circ$;]
 []{bernardi10}. The last two
fields represent possible targets for the LOFAR-EoR observations. Below
we present the main results of these papers.

\begin{figure}
\centering \includegraphics[width=.95\textwidth]{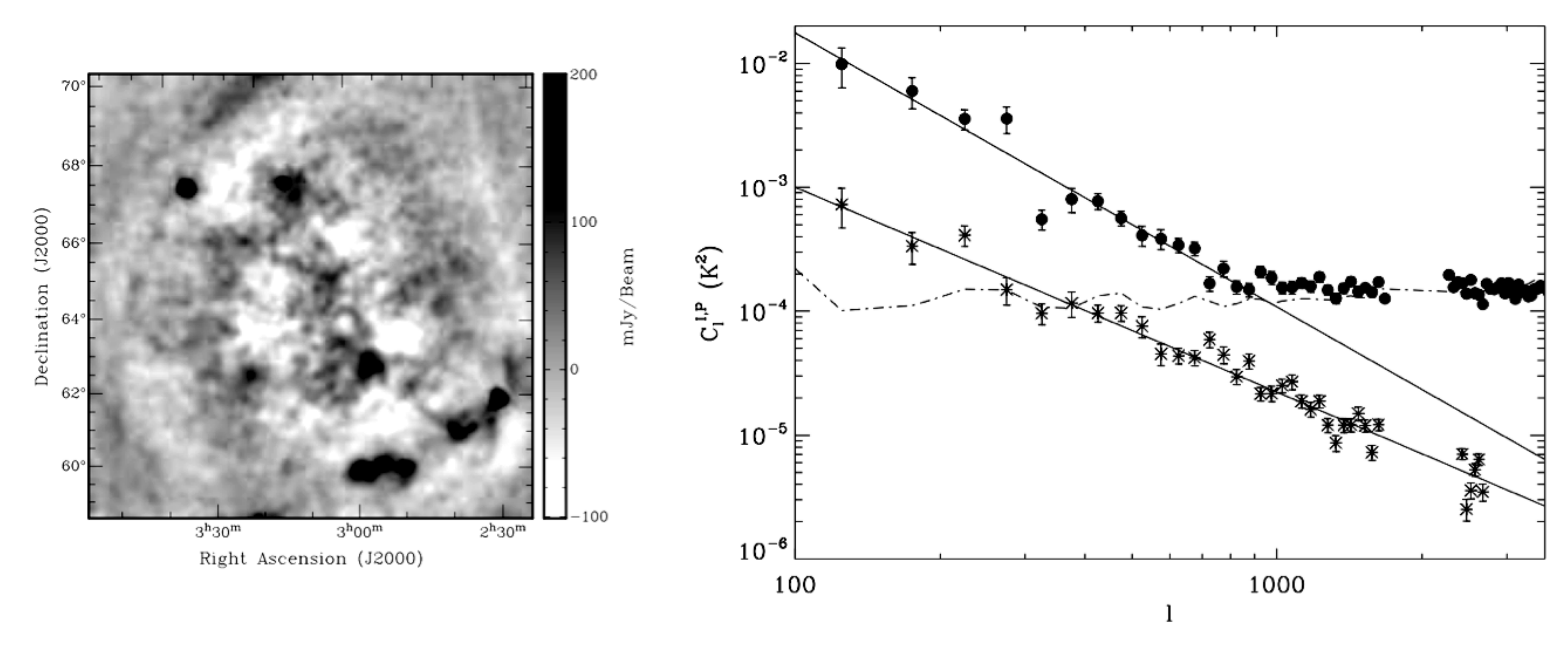}
\caption{Left panel: Stokes I map of the Galactic emission in Fan region. The 
conversion factor is $1~{\rm Jy~beam^{-1}}=105.6~{\rm K}$. Right panel: power 
spectrum (filled circles: total intensity; asterisks: polarized intensity) 
of the Galactic emission in Fan region with the best power-law fit. 
\citep{bernardi09a}}
\label{fig:fan}
\end{figure}

In the ``Fan region'', fluctuations of the Galactic diffuse emission were 
detected at $150~{\rm MHz}$ for the first time (see Fig.~\ref{fig:fan}). 
The fluctuations were detected 
both in total and polarized intensity, with an $rms$ of $14~{\rm K}$ 
($13~{\rm arcmin}$ resolution) and $7.2~{\rm K}$ ($4~{\rm arcmin}$ resolution)
respectively \citep{bernardi09a}. Their spatial structure appeared to
have a power law behavior with a slope of $-2.2\pm0.3$ in total intensity
and $-1.65\pm0.15$ in polarized intensity  (see Fig.~\ref{fig:fan}). 
Note that, due to its strong 
polarized emission, the ``Fan region'' is not a representative part of the
 high Galactic latitude sky.

Fluctuations of the total intensity Galactic diffuse emission in the
``3C196'' and ``NGP'' fields were also observed on scales larger than
$30~{\rm arcmin}$, with an $rms$ of $3.3~{\rm K}$ and $5.5~{\rm K}$
respectively. 

Patchy polarized emission was found in the ``3C196''
field, with an $rms$ value of $0.68~{\rm K}$ on scales larger than
 $30~{\rm arcmin}$ \citep{bernardi10}.  Thus, 
the Galactic polarized emission fluctuations seem to be smaller than
expected by extrapolating from higher frequency
observations. 

\section{Simulations}
The foregrounds in the context of the EoR measurements have been
studied theoretically by various authors. \cite{shaver99} have given
the first overview of the foreground components. \cite{dimatteo02, dimatteo04} 
have studied emission from unresolved
extragalactic sources at low radio frequencies. \cite{oh03} and
\cite{cooray04} have considered the effect of free-free emission from
extragalactic haloes. \cite{santos05} carried out a detailed study of 
the functional form of the foreground correlations. \cite{jelic08} have made 
the first detailed foreground 
model and have simulated the maps that include both the diffuse emission 
from our Galaxy and extragalactic 
sources (radio galaxies and clusters). \cite{gleser08} have also studied both
galactic and extragalactic foregrounds. \cite{wilman08} have developed 
a semi-empirical simulation of the extragalactic radio continuum sky that can be used
as a extragalactic foreground model. \cite{deoliveira08} has used all
publicly available total power radio surveys to obtain 
all-sky Galactic maps at the desired frequency range and 
\cite{sun08, waelkens09, sun09} has developed a detailed Galactic 3D emission
simulation that can be used as a Galactic foreground model. \cite{bowman09} has
studied foreground contamination in the context of the power spectrum estimation. 

\subsection{LOFAR-EoR foreground model}
\begin{figure}
\centering \includegraphics[width=.70\textwidth]{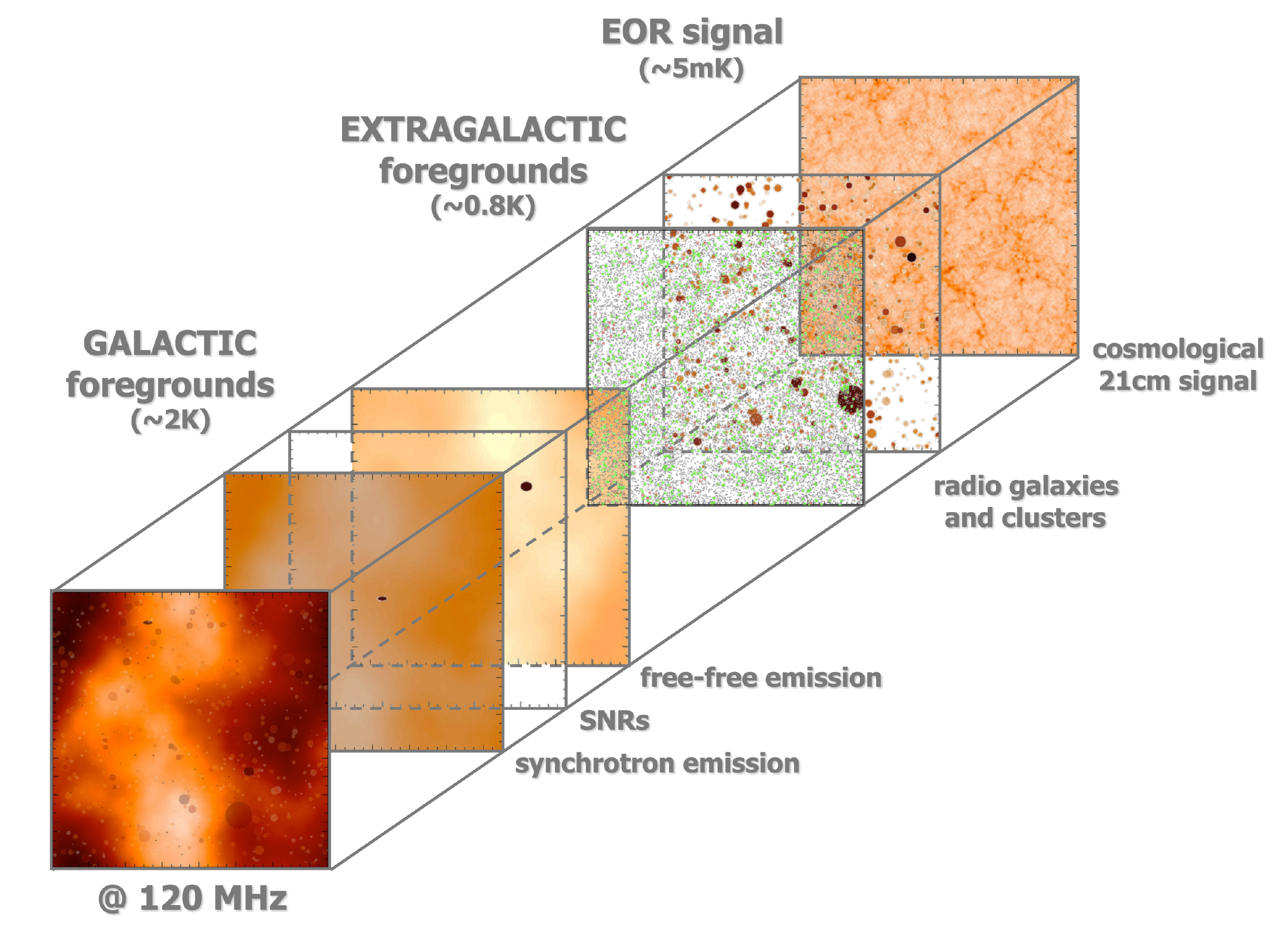}
\caption{The various simulated Galactic and extragalactic contaminants
  of the redshifted 21~cm radiation from the EoR. \citep{jelic08}}
\label{fig:cube}
\end{figure}

Here we give an overview of the foreground model \citep[][and \textit{submitted}]{jelic08}
that is used as a part of the LOFAR-EoR testing pipeline. The
model encompasses the Galactic diffuse synchrotron \& free-free
emission, synchrotron emission from Galactic supernova remnants and
extragalactic emission from radio galaxies and clusters. The simulated 
foreground maps are pertaining in their angular and
frequency characteristics to the LOFAR-EoR experiment (see
Fig.~\ref{fig:cube}).

Since the diffuse Galactic synchrotron emission is the dominant
foreground component, all its observed characteristics are included
in our model: spatial and frequency variations of brightness
temperature and its spectral index, and also the brightness
temperature variations along the line-of-sight. Moreover, the Galactic
emission is derived from physical quantities and the actual
characteristics of our Galaxy (e.g. the cosmic ray and thermal
electron density, and the magnetic field). Thus, the model has the
flexibility to simulate any peculiar case of the Galactic emission
including very complex polarized structures produced by Faraday
screens and depolarization. 

Despite the minor contribution of the Galactic free-free emission, 
it is included in our simulations of the
foregrounds as an individual component. It has a different temperature
spectral index from Galactic synchrotron emission.

Integrated emission from extragalactic sources is decomposed into two
components: emission from radio galaxies and from radio
clusters. Simulations of radio galaxies are based on the source count
functions at low radio frequency by \cite{jackson05}, for three
different types of radio galaxies, namely FR{\small I}, FR{\small II}
and star forming galaxies. Correlations obtained by radio galaxy
surveys are used for their angular distribution. Simulations of radio
clusters are based on a cluster catalogue from the Virgo consortium and
observed mass--X-ray luminosity and X-ray--radio luminosity relations.

\section{Problem of the foreground removal}
\begin{figure}
\centering \includegraphics[width=.95\textwidth]{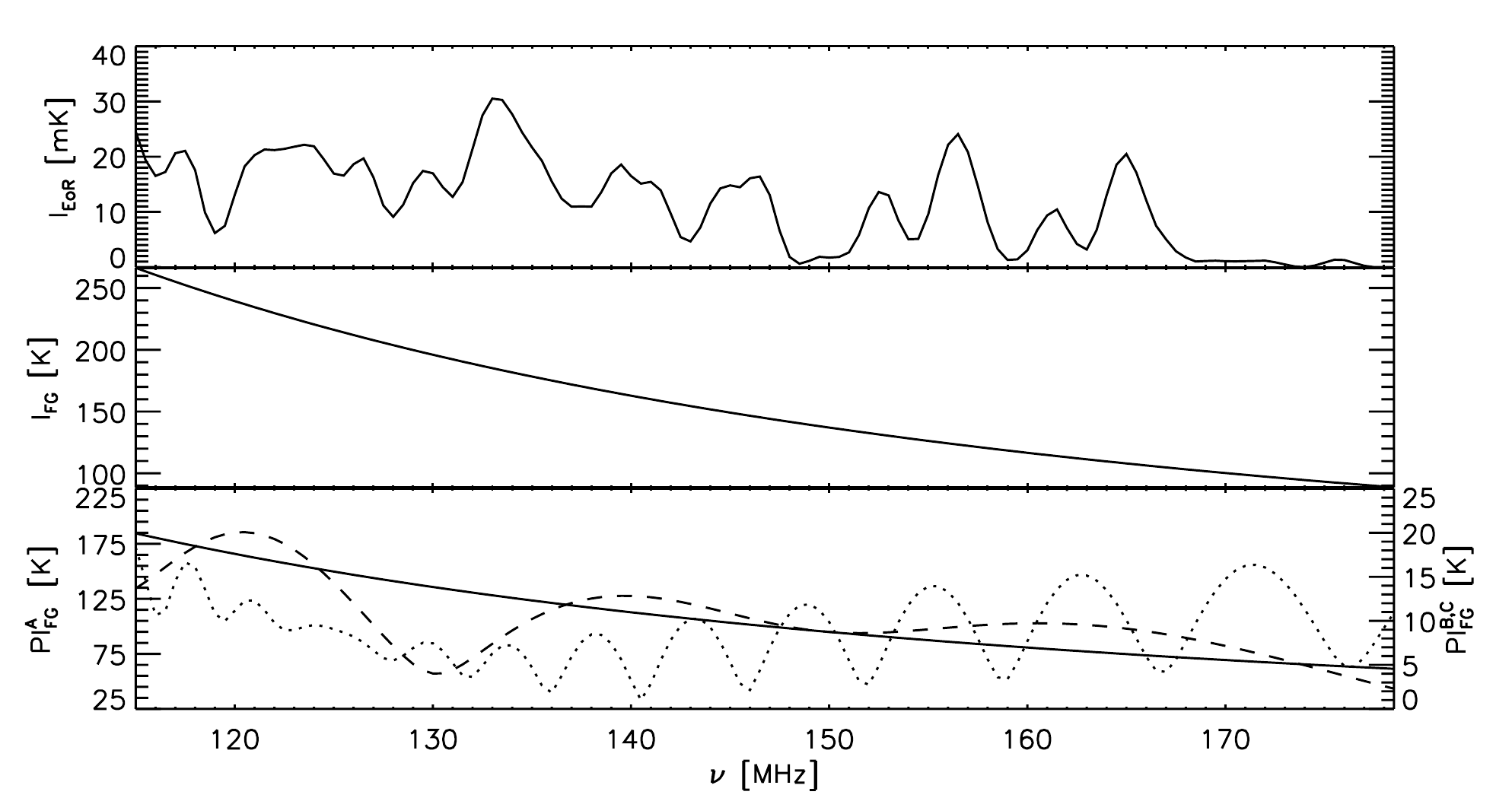}
\caption{A random line through simulated data cubes (images at different frequencies): cosmological 21cm signal in total intensity ($I_{EoR}$,\cite{thomas09}), foregrounds in total intensity ($I_{FG}$), and foregrounds in polarized intensity without ($PI_{FG}^{A}$, solid line) and with differential Faraday rotation in the Galaxy ($PI_{FG}^{B,C}$, dashed and dotted line). Note that the foregrounds are smooth along frequency in total intensity and might show fluctuations in polarized intensity. \cite{jelic10}}
\label{fig:los}
\end{figure}

One of the major challenges of the EoR experiments is the extraction
of the EoR signal from the prominent astrophysical foregrounds. The 
extraction is usually formed in total intensity along frequency, since (see Fig.~\ref{fig:los}):
\begin{enumerate}
\item the cosmological 21~cm signal is essentially unpolarized and fluctuates
along frequency;
\item the foregrounds are smooth along frequency  
in total intensity and might show fluctuations in polarized intensity. 
\end{enumerate}
Thus, the EoR signal can be extracted from the foreground emission 
by fitting out the smooth component of the foregrounds along frequency 
(see Sec.~\ref{sec:meth}). However, all current EoR radio interferometric
arrays have an instrumentally polarized response, 
which needs to be calibrated. If the calibration is  imperfect, some part of the 
polarized signal is transferred into a total intensity and vice versa (hereafter 
`leakage').  As a result, the extraction of the EoR signal is more demanding
(see Sec.~\ref{sec:pol}).

\subsection{LOFAR-EoR experiment: foreground removal methods}\label{sec:meth}
The simplest method for foreground removal in total intensity is a polynomial fitting in
logarithmic scale. We have showed this using the LOFAR-EoR testing pipeline 
\citep{jelic08}. However, one should be careful in choosing the order of the polynomial to
perform the fitting. If the order of the polynomial is too small, the
foregrounds will be under-fitted and the EoR signal could be dominated
and corrupted by the fitting residuals, while if the order of the
polynomial is too big, the EoR signal could be fitted out. Therefore, 
we have argued that in principle it would be better to fit the foregrounds 
non-parametrically (e.g. Wp smoothing) -- allowing the data to determine their shape -- 
rather than selecting some functional form in advance and then fitting its 
parameters \citep{harker09a}. 

\begin{figure}
\centering \includegraphics[width=.95\textwidth]{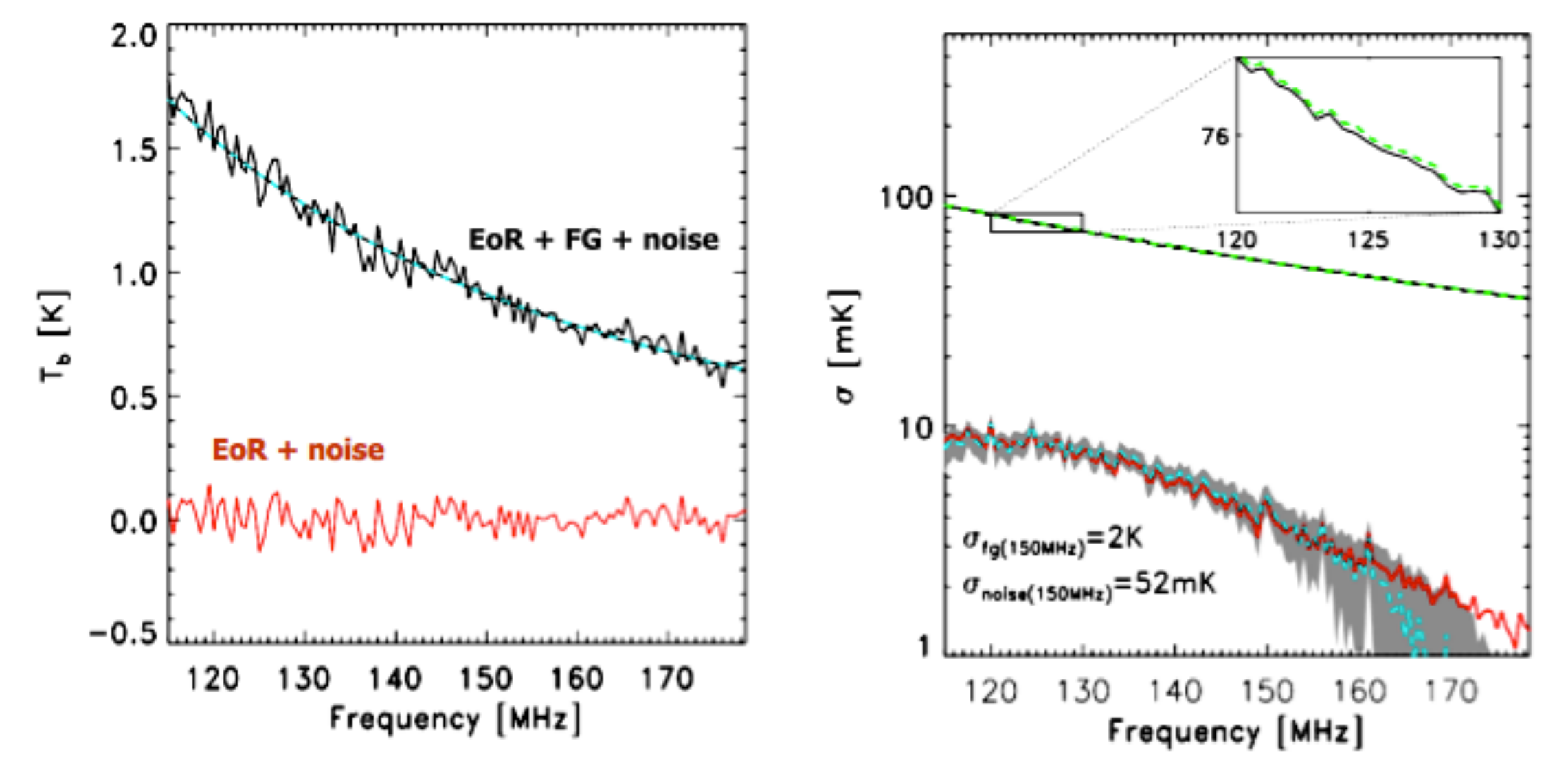}
\caption{\textit{Left panel:} One pixel along frequency of the
simulated LOFAR-EoR data (solid black line), fitted smooth
component of the foregrounds (dashed cyan  line), and 
residuals (solid red line) after taking out the foregrounds. 
Note that the residuals are the sum of the EoR signal and the noise.
\textit{Right panel:} Statistical detection of the EoR signal from the
simulated LOFAR-EoR data, that include diffuse components of the
foregrounds and realistic noise. The dashed-dotted black line 
represents the standard deviation ($\sigma$)
of the noise as a function of frequency, the dotted green line the
$\sigma$ of the residuals after taking out the smooth foreground
component, and the solid red line the $\sigma$ of the original EoR
signal. The grey shaded zone shows the $2\sigma$ detection, whereas
the dashed white line shows the mean of the detection. Note that the
y-axis is in logarithmic scale. \citep{jelic08}}
\label{fig:fgext}
\end{figure}

After foreground subtraction from the EoR observations, 
the residuals will be dominated by instrumental
noise, i.e., the level of the noise is expected to be order of magnitudes 
larger than the EoR signal (assuming 300 h observation with LOFAR). Thus, 
general statistical properties of the noise should be determined and be 
used to statistically detect the cosmological 21 cm signal: e.g., variance of 
the EoR signal over the image,  $\sigma_{\rm EoR}^{2}$, as a 
function of frequency (redshift) obtained by
subtracting the variance of the noise, $\sigma_{\rm noise}^{2}$, 
from that of the residuals, $\sigma_{\rm residuals}^{2}$. We have
showed such statistical detection of the EoR signal using the
fiducial model of the LOFAR-EoR experiment \citep{jelic08} (see Fig.). 
Note that we have also obtained similar results by using different statistics:
the skewness of the one-point distribution of brightness temperature of the
EoR signal, measured as a function of observed frequency \citep{harker09b};
and the power spectrum of variations in the intensity of redshifted 21 cm 
radiation from the EoR \citep{harker10}. 

\subsection{Problem of the `leaked' polarized foregrounds}\label{sec:pol}
\begin{figure}
\centering \includegraphics[width=.95\textwidth]{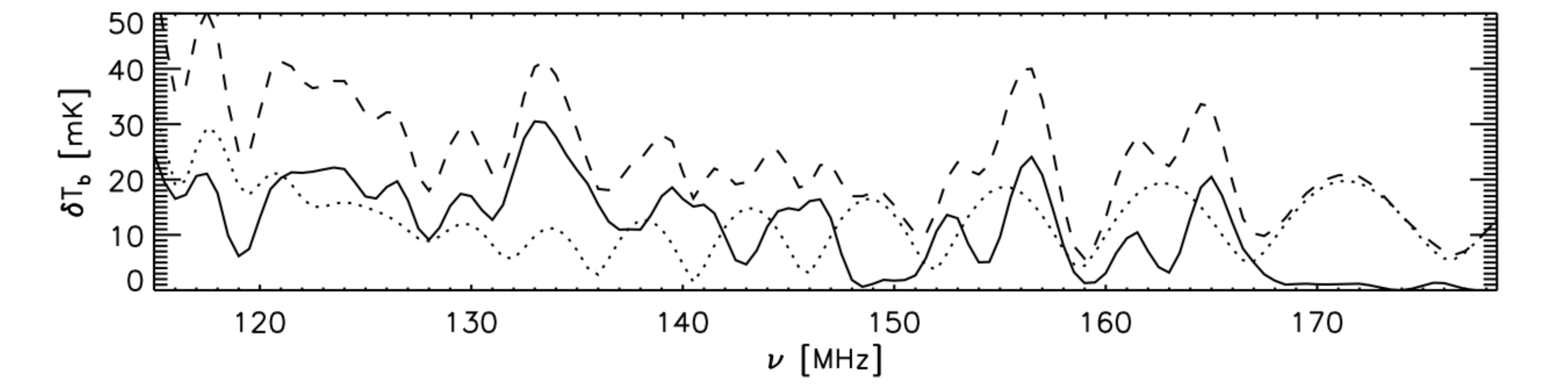}
\caption{A random line of sight through a simulated 21~cm data cube (solid line). Dotted line shows the `leakage' of the polarized Galactic emission to the total intensity and dashed line is a sum of the two. We assume 0.15\%  residual `leakage' and we use $PI_{FG}^C$ (see Fig.~4) as an example of the Galactic emission. The angular and frequency resolution of the data match that of the LOFAR telescope. \cite{jelic10}}
\label{fig:leak}
\end{figure}

In the regions of the Galaxy that have thermal and cosmic-ray electrons 
mixed, that show significant variations of magnetic field, or both, the polarization 
angles of Galactic emission are differentially Faraday rotated. If differential
Faraday rotation is strong enough, the Galactic polarized emission from that region 
can behave along frequency in a manner similar to the cosmological 21 cm signal 
(see Fig.~\ref{fig:los}). Thus, an inaccurate calibration of instrumental polarized response 
can transfer a part of such polarized foreground emission in total intensity and mask
the desired EoR signal (see Fig.~\ref{fig:leak}). 

We have addressed this for the first time through realistic simulations of the 
LOFAR-EoR experiment \citep{jelic10}. Based on our simulations, 
we have concluded that the EoR observational windows need to be in the regions 
with a very low polarized foreground emission, in order to
minimize `leaked' foregrounds. Faraday rotation measure synthesis, 
polarization surveys obtained by different radio telescopes, and a multiple EoR
observations will help in mitigating the polarization leakages. However, further
simulations and observations are necessary to pin point the best strategy for the
EoR detection.

\section{Future perspective}
At the moment, the battle to detect the cosmological 21~cm signal is
fought on two fronts. One in bettering the theoretical understanding
of the Epoch of Reionization and its observational probes, while the
other involves an engineering effort to develop and build next
generation radio telescopes capable of detecting the cosmological
21~cm signal despite a slew of astrophysical and non-astrophysical
contaminants.

The LOFAR-EoR key science project is currently in excellent
shape. Almost all modules of the LOFAR-EoR end-to-end pipeline
\citep{jelicPhD,jelic08,thomas09,offringa10,panos09} are
developed and the pipeline is used intensively for testing the
cosmological signal extraction schemes \citep{jelic08,harker09a,
harker09b,harker10} for the extremely challenging
EoR observations. The observations are carried out using WSRT 
radio telescope to directly measure the properties of the foreground
emission in the frequency range relevant for the EoR experiments
\cite{bernardi09a,bernardi10}.
On the other hand the LOFAR telescope is on
schedule.

The first round of the LOFAR-EoR observations are scheduled for the
end this year. Prior to those observations a shallow survey of the
Northern sky will be preformed. The goal of that survey is to explore
the foreground emission and select the optimal LOFAR-EoR 
observing windows.

The near future will be very interesting and exciting. Observations with LOFAR will provide the deepest images of the low frequency radio sky. Those images, beside the primary aim of probing the EoR, can be used for additional cutting-edge scientific studies. Examples are peculiar cases of Galactic emission in total and
polarized intensity, physics of the Galactic emission processes, properties the Galactic magnetic fields, distribution of the cosmic ray and thermal electrons, source counts and redshift evolution of the
radio galaxies and clusters.

\bibliographystyle{JHEP}
\bibliography{reflist}

\providecommand{\href}[2]{#2}\begingroup\raggedright\begin{thebibliography}{10}

\bibitem{jelicPhD}
V.~{Jeli{\'c}}, {\it {Cosmological 21cm experiments: Searching for a needle in a
  haystack}},  {\em PhD thesis, RuG} (May, 2010)
  http://dissertations.ub.rug.nl/faculties/science/2010/v.jelic/.

\bibitem{jelic08}
V.~{Jeli{\'c}}, S.~{Zaroubi}, P.~{Labropoulos}, R.~M. {Thomas}, G.~{Bernardi},
  M.~A. {Brentjens}, A.~G. {de Bruyn}, B.~{Ciardi}, G.~{Harker}, L.~V.~E.
  {Koopmans}, V.~N. {Pandey}, J.~{Schaye}, and S.~{Yatawatta}, {\it {Foreground
  simulations for the LOFAR-epoch of reionization experiment}},  {\em \mnras}
  {\bf 389} (Sept., 2008) 1319--1335,
  [\href{http://xxx.lanl.gov/abs/0804.1130}{{\tt arXiv:0804.1130}}].

\bibitem{jelic10}
V.~{Jeli{\'c}}, S.~{Zaroubi}, P.~{Labropoulos}, G.~{Bernardi}, A.~G. {de
  Bruyn}, and L.~V.~E. {Koopmans}, {\it {Realistic Simulations of the Galactic
  Polarized Foreground: Consequences for 21-cm Reionization Detection
  Experiments}},  {\em ArXiv e-prints} (July, 2010)
  [\href{http://xxx.lanl.gov/abs/1007.4135}{{\tt arXiv:1007.4135}}].

\bibitem{thomas09}
R.~M. {Thomas}, S.~{Zaroubi}, B.~{Ciardi}, A.~H. {Pawlik}, P.~{Labropoulos},
  V.~{Jeli{\'c}}, G.~{Bernardi}, M.~A. {Brentjens}, A.~G. {de Bruyn}, G.~J.~A.
  {Harker}, L.~V.~E. {Koopmans}, G.~{Mellema}, V.~N. {Pandey}, J.~{Schaye}, and
  S.~{Yatawatta}, {\it {Fast large-scale reionization simulations}},  {\em
  \mnras} {\bf 393} (Feb., 2009) 32--48,
  [\href{http://xxx.lanl.gov/abs/0809.1326}{{\tt arXiv:0809.1326}}].

\bibitem{offringa10}
A.~R. {Offringa}, A.~G. {de Bruyn}, S.~{Zaroubi}, and M.~{Biehl}, {\it {A LOFAR
  RFI detection pipeline and its first results}},  {\em ArXiv e-prints} (July,
  2010) [\href{http://xxx.lanl.gov/abs/1007.2089}{{\tt arXiv:1007.2089}}].

\bibitem{panos09}
P.~{Labropoulos}, L.~V.~E. {Koopmans}, V.~{Jelic}, S.~{Yatawatta}, R.~M.
  {Thomas}, G.~{Bernardi}, M.~{Brentjens}, G.~{de Bruyn}, B.~{Ciardi},
  G.~{Harker}, A.~{Offringa}, V.~N. {Pandey}, J.~{Schaye}, and S.~{Zaroubi},
  {\it {The LOFAR EoR Data Model: (I) Effects of Noise and Instrumental
  Corruptions on the 21-cm Reionization Signal-Extraction Strategy}},  {\em
  ArXiv e-prints} (Jan., 2009) [\href{http://xxx.lanl.gov/abs/0901.3359}{{\tt
  arXiv:0901.3359}}].

\bibitem{shaver99}
P.~A. {Shaver}, R.~A. {Windhorst}, P.~{Madau}, and A.~G. {de Bruyn}, {\it {Can
  the reionization epoch be detected as a global signature in the cosmic
  background?}},  {\em \aap} {\bf 345} (May, 1999) 380--390,
  [\href{http://xxx.lanl.gov/abs/astro-ph/9901320}{{\tt astro-ph/9901320}}].

\bibitem{landecker70}
T.~L. {Landecker} and R.~{Wielebinski}, {\it {The Galactic Metre Wave
  Radiation: A two-frequency survey between declinations $+25^{o}$ and
  $-25^{o}$ and the preparation of a map of the whole sky}},  {\em Australian
  Journal of Physics Astrophysical Supplement} {\bf 16} (1970) 1--+.

\bibitem{ali08}
S.~S. {Ali}, S.~{Bharadwaj}, and J.~N. {Chengalur}, {\it {Foregrounds for
  redshifted 21-cm studies of reionization: Giant Meter Wave Radio Telescope
  153-MHz observations}},  {\em \mnras} {\bf 385} (Apr., 2008) 2166--2174,
  [\href{http://xxx.lanl.gov/abs/0801.2424}{{\tt arXiv:0801.2424}}].

\bibitem{rogers08}
A.~E.~E. {Rogers} and J.~D. {Bowman}, {\it {Spectral Index of the Diffuse Radio
  Background Measured from 100 TO 200 MHz}},  {\em \aj} {\bf 136} (Aug., 2008)
  641--648, [\href{http://xxx.lanl.gov/abs/0806.2868}{{\tt arXiv:0806.2868}}].

\bibitem{pen09}
U.-L. {Pen}, T.-C. {Chang}, C.~M. {Hirata}, J.~B. {Peterson}, J.~{Roy},
  Y.~{Gupta}, J.~{Odegova}, and K.~{Sigurdson}, {\it {The GMRT EoR experiment:
  limits on polarized sky brightness at 150 MHz}},  {\em \mnras} (Sept., 2009)
  1240--+, [\href{http://xxx.lanl.gov/abs/0807.1056}{{\tt arXiv:0807.1056}}].

\bibitem{bernardi09a}
G.~{Bernardi}, A.~G. {de Bruyn}, M.~A. {Brentjens}, B.~{Ciardi}, G.~{Harker},
  V.~{Jeli{\'c}}, L.~V.~E. {Koopmans}, P.~{Labropoulos}, A.~{Offringa}, V.~N.
  {Pandey}, J.~{Schaye}, R.~M. {Thomas}, S.~{Yatawatta}, and S.~{Zaroubi}, {\it
  {Foregrounds for observations of the cosmological 21 cm line. I. First
  Westerbork measurements of Galactic emission at 150 MHz in a low latitude
  field}},  {\em \aap} {\bf 500} (June, 2009) 965--979,
  [\href{http://xxx.lanl.gov/abs/0904.0404}{{\tt arXiv:0904.0404}}].

\bibitem{bernardi10}
G.~{Bernardi}, A.~G.~d.~B.~G. {Harker}, M.~A. {Brentjens}, B.~{Ciardi},
  V.~{Jeli{\'c}}, L.~V.~E. {Koopmans}, P.~{Labropoulos}, A.~{Offringa}, V.~N.
  {Pandey}, J.~{Schaye}, R.~M. {Thomas}, S.~{Yatawatta}, and S.~{Zaroubi}, {\it
  {Foregrounds for observations of the cosmological 21 cm line: II. Westerbork
  observations of the fields around 3C196 and the North Celestial Pole}},  {\em
  ArXiv e-prints} (Feb., 2010) [\href{http://xxx.lanl.gov/abs/1002.4177}{{\tt
  arXiv:1002.4177}}].

\bibitem{laing83}
R.~A. {Laing}, J.~M. {Riley}, and M.~S. {Longair}, {\it {Bright radio sources
  at 178 MHz - Flux densities, optical identifications and the cosmological
  evolution of powerful radio galaxies}},  {\em \mnras} {\bf 204} (July, 1983)
  151--187.

\bibitem{hales88}
S.~E.~G. {Hales}, J.~E. {Baldwin}, and P.~J. {Warner}, {\it {The 6C survey of
  radio sources. II - The zone delta = 30-51 deg, alpha = 08h30m-17h30m}},
  {\em \mnras} {\bf 234} (Oct., 1988) 919--936.

\bibitem{debruyn98}
G.~{de Bruyn}, G.~{Miley}, R.~{Rengelink}, and {et al.}, {\em {WENSS}}.
\newblock ASTRON, 1998.

\bibitem{reich06}
W.~{Reich}, {\it {Galactic polarization surveys}},  {\em ArXiv Astrophysics
  e-prints} (Mar., 2006) [\href{http://xxx.lanl.gov/abs/astro-ph/0603465}{{\tt
  astro-ph/0603465}}].

\bibitem{dimatteo02}
T.~{Di Matteo}, R.~{Perna}, T.~{Abel}, and M.~J. {Rees}, {\it {Radio
  Foregrounds for the 21 Centimeter Tomography of the Neutral Intergalactic
  Medium at High Redshifts}},  {\em \apj} {\bf 564} (Jan., 2002) 576--580,
  [\href{http://xxx.lanl.gov/abs/astro-ph/0109241}{{\tt astro-ph/0109241}}].

\bibitem{dimatteo04}
T.~{Di Matteo}, B.~{Ciardi}, and F.~{Miniati}, {\it {The 21-cm emission from
  the reionization epoch: extended and point source foregrounds}},  {\em
  \mnras} {\bf 355} (Dec., 2004) 1053--1065,
  [\href{http://xxx.lanl.gov/abs/astro-ph/0402322}{{\tt astro-ph/0402322}}].

\bibitem{oh03}
S.~P. {Oh} and K.~J. {Mack}, {\it {Foregrounds for 21-cm observations of
  neutral gas at high redshift}},  {\em \mnras} {\bf 346} (Dec., 2003)
  871--877, [\href{http://xxx.lanl.gov/abs/astro-ph/0302099}{{\tt
  astro-ph/0302099}}].

\bibitem{cooray04}
A.~{Cooray}, {\it {Cross-correlation studies between CMB temperature
  anisotropies and 21cm fluctuations}},  {\em \prd} {\bf 70} (Sept., 2004)
  063509--+, [\href{http://xxx.lanl.gov/abs/astro-ph/0405528}{{\tt
  astro-ph/0405528}}].

\bibitem{santos05}
M.~G. {Santos}, A.~{Cooray}, and L.~{Knox}, {\it {Multifrequency Analysis of 21
  Centimeter Fluctuations from the Era of Reionization}},  {\em \apj} {\bf 625}
  (June, 2005) 575--587, [\href{http://xxx.lanl.gov/abs/astro-ph/0408515}{{\tt
  astro-ph/0408515}}].

\bibitem{gleser08}
L.~{Gleser}, A.~{Nusser}, and A.~J. {Benson}, {\it {Decontamination of
  cosmological 21-cm maps}},  {\em \mnras} {\bf 391} (Nov., 2008) 383--398,
  [\href{http://xxx.lanl.gov/abs/0712.0497}{{\tt arXiv:0712.0497}}].

\bibitem{wilman08}
R.~J. {Wilman}, L.~{Miller}, M.~J. {Jarvis}, T.~{Mauch}, F.~{Levrier}, F.~B.
  {Abdalla}, S.~{Rawlings}, H.~{Kl{\"o}ckner}, D.~{Obreschkow}, D.~{Olteanu},
  and S.~{Young}, {\it {A semi-empirical simulation of the extragalactic radio
  continuum sky for next generation radio telescopes}},  {\em \mnras} {\bf 388}
  (Aug., 2008) 1335--1348, [\href{http://xxx.lanl.gov/abs/0805.3413}{{\tt
  arXiv:0805.3413}}].

\bibitem{deoliveira08}
A.~{de Oliveira-Costa}, M.~{Tegmark}, B.~M. {Gaensler}, J.~{Jonas}, T.~L.
  {Landecker}, and P.~{Reich}, {\it {A model of diffuse Galactic radio emission
  from 10 MHz to 100 GHz}},  {\em \mnras} {\bf 388} (July, 2008) 247--260,
  [\href{http://xxx.lanl.gov/abs/0802.1525}{{\tt arXiv:0802.1525}}].

\bibitem{sun08}
X.~H. {Sun}, W.~{Reich}, A.~{Waelkens}, and T.~A. {En{\ss}lin}, {\it {Radio
  observational constraints on Galactic 3D-emission models}},  {\em \aap} {\bf
  477} (Jan., 2008) 573--592, [\href{http://xxx.lanl.gov/abs/0711.1572}{{\tt
  arXiv:0711.1572}}].

\bibitem{waelkens09}
A.~{Waelkens}, T.~{Jaffe}, M.~{Reinecke}, F.~S. {Kitaura}, and T.~A.
  {En{\ss}lin}, {\it {Simulating polarized Galactic synchrotron emission at all
  frequencies. The Hammurabi code}},  {\em \aap} {\bf 495} (Feb., 2009)
  697--706, [\href{http://xxx.lanl.gov/abs/0807.2262}{{\tt arXiv:0807.2262}}].

\bibitem{sun09}
X.~H. {Sun} and W.~{Reich}, {\it {Simulated SKA maps from Galactic 3D-emission
  models}},  {\em ArXiv e-prints} (Aug., 2009)
  [\href{http://xxx.lanl.gov/abs/0908.3378}{{\tt arXiv:0908.3378}}].

\bibitem{bowman09}
J.~D. {Bowman}, M.~F. {Morales}, and J.~N. {Hewitt}, {\it {Foreground
  Contamination in Interferometric Measurements of the Redshifted 21 cm Power
  Spectrum}},  {\em \apj} {\bf 695} (Apr., 2009) 183--199,
  [\href{http://xxx.lanl.gov/abs/0807.3956}{{\tt arXiv:0807.3956}}].

\bibitem{jackson05}
C.~{Jackson}, {\it {The Extragalactic Radio Sky at Faint Flux Densities}},
  {\em Publications of the Astronomical Society of Australia} {\bf 22} (2005)
  36--48.

\bibitem{harker09a}
G.~{Harker}, S.~{Zaroubi}, G.~{Bernardi}, M.~A. {Brentjens}, A.~G. {de Bruyn},
  B.~{Ciardi}, V.~{Jeli{\'c}}, L.~V.~E. {Koopmans}, P.~{Labropoulos},
  G.~{Mellema}, A.~{Offringa}, V.~N. {Pandey}, J.~{Schaye}, R.~M. {Thomas}, and
  S.~{Yatawatta}, {\it {Non-parametric foreground subtraction for 21-cm epoch
  of reionization experiments}},  {\em \mnras} {\bf 397} (Aug., 2009)
  1138--1152, [\href{http://xxx.lanl.gov/abs/0903.2760}{{\tt
  arXiv:0903.2760}}].

\bibitem{harker09b}
G.~J.~A. {Harker}, S.~{Zaroubi}, R.~M. {Thomas}, V.~{Jeli{\'c}},
  P.~{Labropoulos}, G.~{Mellema}, I.~T. {Iliev}, G.~{Bernardi}, M.~A.
  {Brentjens}, A.~G. {de Bruyn}, B.~{Ciardi}, L.~V.~E. {Koopmans}, V.~N.
  {Pandey}, A.~H. {Pawlik}, J.~{Schaye}, and S.~{Yatawatta}, {\it {Detection
  and extraction of signals from the epoch of reionization using higher-order
  one-point statistics}},  {\em \mnras} {\bf 393} (Mar., 2009) 1449--1458,
  [\href{http://xxx.lanl.gov/abs/0809.2428}{{\tt arXiv:0809.2428}}].

\bibitem{harker10}
G.~{Harker}, S.~{Zaroubi}, G.~{Bernardi}, M.~A. {Brentjens}, A.~G. {de Bruyn},
  B.~{Ciardi}, V.~{Jeli{\'c}}, L.~V.~E. {Koopmans}, P.~{Labropoulos},
  G.~{Mellema}, A.~{Offringa}, V.~N. {Pandey}, A.~H. {Pawlik}, J.~{Schaye},
  R.~M. {Thomas}, and S.~{Yatawatta}, {\it {Power spectrum extraction for
  redshifted 21-cm Epoch of Reionization experiments: the LOFAR case}},  {\em
  \mnras} {\bf 405} (July, 2010) 2492--2504,
  [\href{http://xxx.lanl.gov/abs/1003.0965}{{\tt arXiv:1003.0965}}].

\end{thebibliography}\endgroup

\end{document}